\documentstyle[]{aipproc}

\begin{document}

\title{Mergers and Galaxy Assembly}

\author{Joshua E. Barnes}

\address{Institute for Astronomy, University of Hawai`i\\
         2680 Woodlawn Drive, Honolulu, Hawai`i, 96822, USA}

\maketitle

\begin{abstract}
Theoretical considerations and observational data support the idea
that mergers were more frequent in the past.  At redshifts $z = 2$
to~$5$, violent interactions and mergers may be implicated by
observations of Lyman-break galaxies, sub-mm starbursts, and active
galactic nuclei.  Most stars in cluster ellipticals probably formed at
such redshifts, as did most of the halo and globular clusters of the
Milky Way; these events may all be connected with mergers.  But what
{\it kind\/} of galaxies merged at high redshifts, and are
present-epoch mergers useful guides to these early collisions?  I will
approach these questions by describing ideas for the formation of the
Milky Way, elliptical galaxies, and systems of globular clusters.
\end{abstract}

\section*{INTRODUCTION}

Why is it so plausible that galactic mergers and tidal interactions
were more frequent in the past?  Several theoretical reasons come to
mind:

\begin{itemize}

\item
Hierarchical clustering, in which small objects are progressively
incorporated into larger structures \cite{L54}, is common to many
accounts of galaxy formation.  In the ``core-halo'' picture
\cite{WR78}, clustering of dark matter creates galaxy halos which
subsequently accumulate cores of baryons, forming visible galaxies.

\item
Tidal encounters generate short-lived features; a population of binary
galaxies with highly eccentric orbits is required to explain the
peculiar galaxies observed today \cite{TT72}.  If these binaries have
a flat distribution of binding energies, their merger rate has
declined with time as $t^{-5/3}$, and the $10$ or so merging galaxies
in the NGC catalog are but the most recent additions to a population
of about $750$ remnants \cite{T77}.

\item
The CDM model \cite{BFPR84} provides a concrete example of galaxy
formation in which merging of dark halos is easily calculated and
clearly important \cite{LC93}.

\end{itemize}

Observations, though not always reaching the redshift range emphasized
in this meeting, also imply rapid merging at high redshift:

\begin{itemize}

\item
Various counting strategies indicate that the pair density grows like
$(1 + z)^m$, where $m \simeq 3 \pm 1$ \cite{ZK89,A99}.

\item
Peculiar morphology becomes more common with increasing redshift
\cite{vdBAETSG96}.  For example, the fraction of irregular galaxies in
the CFRS survey increases from about $10$\% at $z \sim 0.4$ to a third
at $z \sim 0.8$ \cite{BASTELFGHCCB98}.

\end{itemize}

Thus, both theory and observation support the notion that there was
``a great deal of merging of sizable bits and pieces (including quite
a few lesser galaxies) early in the career of every major galaxy''
\cite{T77}.  But the {\it nature\/} of these early mergers is not so
clear; were the objects involved dominated by dark matter, by gas, or
by stars?  And can we learn anything about early mergers by studying
present-epoch examples?

\section*{SIGNPOSTS OF HIGH-REDSHIFT MERGERS}

Merging is hard to prove at redshifts $z \gtrsim 1.5$; cosmological
dimming renders tidal tails nearly invisible, while bandshifting
effects complicate interpretation of the observations \cite{HV97}.
But circumstantial evidence implicates merging in various high-$z$
objects.

\subsection*{Starburst Galaxies}

The most extensive and unbiased sample of high-redshift galaxies are
the ``Lyman-break'' objects at $z \sim 3$, which have rest-frame UV
luminosities consistent with star formation rates of $\sim 10^1
{\rm\,M_\odot\,yr^{-1}}$ \cite{SGPDA96}.  The actual rates could be
several times higher, since much of the UV emitted by young stars may
be absorbed by dust (eg. \cite{H98}).  Spectra show gas outflows with
velocities of $\sim 500 {\rm\,km\,sec^{-1}}$ \cite{PKSDAG98}, atypical
of quiescent galaxies but fairly normal for starburst systems.
Heavily obscured high-$z$ starbursts have been detected at sub-mm
wavelengths \cite{HSDRRBMIPEGOLLGJ98,BCSFTSKO98}.  These have IR
spectral energy distributions similar to ultra-luminous starburst
galaxies like Arp~220 and appear to be forming stars at rates of $\sim
10^2 {\rm\,M_\odot\,yr^{-1}}$.

At low redshifts, luminous starbursts are often triggered by mergers
of gas-rich galaxies \cite{SM96}.  The gas in such systems is highly
concentrated; H$_2$ surface densities of $10^3$ to~$10^5
{\rm\,M_\odot\,pc^{-2}}$ are typical of nearby starbursts \cite{K98},
and similar surface densities are indicated in high-$z$ starbursts
\cite{H98}.  In the potential of an axisymmetric galaxy, gas becomes
``hung up'' in a disk several kpc in radius (Frenk, these proceedings)
instead of flowing inward.  Violently changing potentials in merging
galaxies enable gas to shed its angular momentum and collapse to as
little as $\sim 1$\% of its initial radius \cite{BH96}.

But models based on mergers of low-$z$ disk galaxies may not apply to
high-redshift starbursts \cite{SPF98}.  First, bar instabilities in
isolated galaxies can drive rapid gas inflows without external
triggers \cite{S81}.  Second, disks forming at higher redshifts are
more compact \cite{MMW98} and thus may already have the surface
densities associated with starbursts.  Third, the starbursts in
Lyman-break galaxies occur on scales of several kpc (Weedman, these
proceedings), whereas inflows concentrate gas into much smaller
regions.  Nonetheless, these objects also have irregular morphologies
suggestive of mergers, and deep HDF images reveal faint asymmetric
features which may be due to tidal interactions
\cite{vdBAETSG96,SGDA96}.  Mergers seem to be the ``best bet'' for
high-$z$ starbursts, but something more than naive extrapolation from
low-$z$ is needed to test this conjecture.

\subsection*{Radio Galaxies}

At low redshifts, powerful radio sources are often associated with
merger remnants; some $30$\% exhibit tails, fans, shells, or other
signatures of recent collisions \cite{HSBvBMIBB86}.  But at redshifts
$z \gtrsim 0.6$ the most striking morphological feature of powerful
radio sources is a near-ubiquitous alignment between the radio lobes
and continuum optical emission \cite{MvBSD87,CMvB87}.  This
``alignment effect'' seems at odds with the merger morphologies seen
at low redshift; one explanation invokes jet-induced star formation
(eg. \cite{MvBSD87}).

Recent observations suggest the alignment effect is compatible with
mergers \cite{S99}.  Strong polarization is found in several $z
\gtrsim 2$ radio galaxies, implying that the aligned emission is
scattered light from an obscured AGN (eg. \cite{TFdSA89}); in several
cases there is good evidence that dust is the primary scattering agent
\cite{KC97,RMAP97}.  HST imaging of the radio galaxy 0406--244 at $z =
2.44$ reveals a double nucleus and what appear to be tidal debris
illuminated by an AGN \cite{RMAP97}.

From a theoretical perspective, merging may even be {\it necessary\/}
to form powerful radio sources.  The most plausible engines for such
galaxies are rapidly spinning black holes (Blandford, these
proceedings).  Accretion from a disk can't spin up a black hole unless
the accretion phase lasts $\sim 0.1 {\rm\,Gyr}$; on the other hand,
two black holes of comparable mass can coalesce to produce a
rapidly-spinning hole \cite{WC95}.

\subsection*{Quasars}

Evidence that low-redshift quasars frequently occur in interacting
systems has been accumulating for two decades \cite{S99}.  Early
claims that quasars have close companions are supported by recent
studies out to redshifts $z \sim 1$ \cite{DBBBCDMMSP95,FBKS96,SR98}.
Even more telling are the tidal tails and other signs of violent
interactions in nearby cases \cite{SM83,SR91,BKS95,BDBBCDMMS96,SCC98}.

The very nature of these interactions makes their detection difficult
at higher redshifts -- tidal tails and other signs are hidden by
cosmological dimming and quasar glare.  Nor does the low-$z$ evidence
preclude the possibility that high-redshift quasars may have nothing
to do with mergers.  However, the peak in quasar activity at $z \sim
2$ to~$3$ seems to broadly coincide with other indications of
extensive merging activity reviewed above.  Given the observational
difficulties, a compelling case that this high-$z$ activity is driven
by mergers probably awaits a theory for the formation of supermassive
black holes.

\section*{ASSEMBLING THE MILKY WAY}

Complementing the data gathered by looking back to high redshift is
information gleaned by ``archeological'' studies of objects at $z \sim
0$.  The oldest components of the Milky Way provide evidence that
mergers of small galaxies played an important role \cite{SZ78}:

\begin{enumerate}

\item
A ``second parameter'' -- which may not be age \cite{SvdBB96} -- is
required to account for variations in the stellar content of globular
clusters.

\item
This second parameter is correlated with orbital direction; clusters
with retrograde orbits have Oosterhoff class I variables \cite{vdB93}.

\item
Halo stars with $[{\rm Fe}/{\rm H}] \sim -1$ have a large range of
$[\alpha/{\rm Fe}]$ values \cite{GW98,S98}.

\item
The outer halo exhibits retrograde rotation with respect to the rest
of the galaxy \cite{M96}.

\item
The halo is not completely well-mixed, as indicated by observations of
star streams and moving groups \cite{M96,E87,LBLB95}.

\end{enumerate}

Items 1--3 indicate that different parts of the halo have different
enrichment histories, items 2 \& 4 imply that some part of the halo
fell in on a retrograde orbit, and item 5 is direct evidence for the
gradual dissolution of fragments after merging.

Halo accretion is clearly an ongoing process, as shown by the
discovery of the Sgr~I dwarf galaxy \cite{IGI95} and by observations
of high-latitude A stars \cite{PBS94}.  But two different arguments
suggest that the {\it bulk\/} of the halo fell into place long ago.

First, halo stars are old.  The halo as a whole shows a well-defined
turn-off at $B - V \sim 0.4$, corresponding to ages $\gtrsim 10
{\rm\,Gyr}$; only $\sim 10$\% of the stars appear younger
\cite{UWG96}.  To be sure, this does not rule out recent accretions of
objects containing only old stars, but most dwarf galaxies in the
local group contain intermediate-age stars as well.  Thus, unless the
accreted galaxies were unlike those we observe today, most fell in
more than $10 {\rm\,Gyr}$ ago.

Second, galactic disks are dynamically fragile; accretion of satellite
galaxies can easily ruin a stellar disk.  Analytic estimates limit the
mass accreted by the Milky Way to less than $4$\% in the past $5
{\rm\,Gyr}$ \cite{TO92}.  N-body experiments show less disk heating
than the analytic work predicts; dark halos absorb much of the damage,
and disks may tilt as well as thicken \cite{WMH96,HC97,VW98}.  Still,
accretion events of any size increase the disk's vertical dispersion,
$\sigma_{\rm z}$.  Significant structure is seen in the $\sigma_{\rm
z}$--age relation; most striking is the jump from $\sigma_{\rm z}
\simeq 20$ to~$40 {\rm\,km\,sec^{-1}}$ which marks the transition to
the $\sim 10 {\rm\,Gyr}$-old thick disk \cite{F93}.

In sum, the Milky Way last suffered a significant merger at least $10
{\rm\,Gyr}$ ago; relics of this event include the outer stellar halo
and possibly the thick disk.  Presumably, the Milky Way's dark halo
was largely in place at this time, since a major merger would have
disrupted even the thick disk.

\section*{ASSEMBLING CLUSTER ELLIPTICALS}

Galaxy clusters are old in two distinct respects: first, cluster
galaxies probably collapsed early; second, dynamical processes run
faster in proportion to $\sqrt{\rho}$.  Thus clusters should contain
remnants of many high-redshift mergers.  Archeological evidence from
nearby clusters provides important clues to these mergers.

\subsection*{Merger Formation}

After some controversy, it's generally accepted that elliptical
galaxies can be formed by fairly {\it recent\/} mergers of disk
galaxies.  Support for this position includes:

\begin{itemize}

\item
Studies of proto-elliptical merger remnants like NGC~7252 \cite{Sc82}
and models of disk galaxy mergers reproducing such objects
\cite{B88,HM95}.

\item
H$\beta$ line strengths in some ellipticals indicating recent star
formation \cite{FTGW94}.

\item
``Fine structures'' in elliptical galaxies correlating with residuals
in luminosity--color and luminosity--line strength relations
\cite{SSFBDOCG90,SS92}.

\end{itemize}

These results enable us to trace the gradual assimilation of recent
merger remnants into the larger population of field ellipticals.  But
such evidence is not available for cluster ellipticals, which seem to
be a more homogeneous population (eg. \cite{dCD92}).  Studies of the
fundamental plane out to $z \simeq 0.8$ indicate that cluster
ellipticals evolve passively and probably formed the bulk of their
stars at $z \gtrsim 2$ \cite{vDFKI98}.  Thus cluster ellipticals are
unlikely to show the signs which betray aging merger remnants in the
field.

Counter-rotating or otherwise decoupled ``cores'' are probably the
clearest signs that cluster ellipticals were formed by ancient mergers
\cite{SB95,MSBW98}.  High-resolution imaging shows that kinematically
distinct nuclear components are usually {\it disks\/}
\cite{SB95,CFIF97}.  Such disks typically have high metal abundances
\cite{BS92} and low velocity dispersions \cite{RW92}.  These
properties indicate that they formed dissipationally during major
mergers \cite{FI88,S90}; merger simulations producing counter-rotating
nuclear gas disks back up this hypothesis \cite{HB91}.

The nature of the mergers which formed cluster ellipticals is unknown;
often invoked are highly dissipative encounters of gaseous fragments.
But the existence of counter-rotating disks indicates that the
penultimate participants can't have been very numerous or very gassy.
If many small objects coalesced, the law of averages would make
counter-rotation extremely rare.  And counter-rotation is unlikely to
arise in essentially gaseous mergers since gas flows can't
interpenetrate.

Once formed, kinematically distinct disks would be easily disrupted by
dissipationless mergers \cite{Sc98}.  Thus observations of such
structures in cluster galaxies imply that few mergers occur once a
cluster has virialized.  This is entirely plausible on dynamical
grounds since encounters at speeds higher than about twice a galaxy's
internal velocity dispersion don't result in mergers \cite{MH97}.

\subsection*{Abundance Ratios}

In elliptical galaxies, $\alpha$-process elements are more abundant
with respect to Fe than they are in the disk of the Milky Way
\cite{WFG92}.  This may constrain the timescale for star formation in
ellipticals, since $\alpha$-process elements are produced in SN~II,
which explode on a short timescale, while Fe is also produced in
SN~Ia, which explode after $\sim 1 {\rm\,Gyr}$.  Indeed, $[{\rm
Mg}/{\rm Fe}] \simeq 0.5$ for the nuclear disks in cluster ellipticals
\cite{SB95,MSBW98}.  High $\alpha$-process abundances indicate that
SN~Ia played little role in enriching these galaxies; on the face of
it, they also imply that cluster ellipticals formed on timescales
$\lesssim 1 {\rm\,Gyr}$ (eg. \cite{B97}).

High abundances of $\alpha$-process elements with respect to Fe are
also seen in X-ray observations of the diffuse gas in galaxy clusters
(eg. \cite{MLATFMKH96}, but see \cite{IA97}).  The large amounts of
metals in cluster gas require remarkably high SN rates which may not
be possible with a Salpeter IMF \cite{RCDEP93}.  These results
undermine the argument that high $\alpha/{\rm Fe}$ ratios imply short
enrichment timescales, since abundances in the cluster gas presumably
represent integrated metal production over $\sim 10 {\rm\,Gyr}$.  The
abundance patterns of cluster ellipticals are clearly inconsistent
with mergers of present-day spirals, but do not preclude mergers of
moderately gas-rich galaxies containing substantial stellar disks.

\subsection*{Globular Clusters}

Young star clusters are observed in star-forming galaxies like the LMC
\cite{EF85} and in intense starburst galaxies \cite{MHLKRG95,WS95}.
These clusters have half-light radii of less than $5 {\rm\,pc}$,
masses of $10^4$ to~$10^7 {\rm\,M_\odot}$, and metal abundances
comparable to their parent starbursts.  Their luminosity functions
follow power laws with slopes of $-1.6$ to~$-2$, intriguingly close to
the mass function of giant molecular clouds \cite{HP94}.  However,
it's not entirely clear that cluster luminosity is a good indicator of
mass since some range of cluster ages is usually present.

Evidence is accumulating that the globular cluster systems of field
ellipticals are partly due to cluster formation in merger-induced
starbursts:

\begin{itemize}

\item
Ongoing and recent mergers (eg., NGC~4038/9, NGC~7252, NGC~3921) have
populations of blue luminous clusters with ages of less than $1
{\rm\,Gyr}$ \cite{WS95,SMWF96,MWSF97}.

\item
Older remnants (eg., NGC~3610) have redder and fainter clusters with
ages of a few Gyr \cite{WMSF97}.

\item
Predicted specific frequencies\footnote{The specific frequency $S_{\rm
N}$ is defined as the number of globular clusters divided by the
galaxy luminosity in units of $M_{\rm V} = -15$.} in merger remnants
increase to $S_{\rm N} \simeq 2$ or~$3$ over $\sim 10 {\rm\,Gyr}$ as
the stellar populations fade \cite{SMWF96,MWSF97}.

\item
Globulars in elliptical galaxies have bimodal color (metalicity)
distributions.

\end{itemize}

These findings imply that metal-rich star clusters form during mergers
and are gradually assimilated into existing globular cluster
populations \cite{AZ92}.  However, the large populations of metal-{\it
poor\/} globulars found in cluster ellipticals are {\it not\/}
consistent with mergers of field spirals \cite{FBG97}; predicted
specific frequencies of metal-poor clusters are $S_{\rm N}^{\rm P}
\simeq 1$, while in fact $S_{\rm N}^{\rm P} \simeq 4$.  This problem
is even worse for cluster systems in cD galaxies, which have $S_{\rm
N}^{\rm P} \simeq 10$; obviously, no amount of merging between
metal-rich systems will produce metal-poor clusters!

The question of high-$S_{\rm N}$ in cluster ellipticals boils down to
this: fewer stars, or more globulars?  One way to get fewer stars is
to merge galaxies {\it after\/} their metal-poor globulars have formed
but before they build up substantial disks.  For example, the Milky
Way as it was $\sim 10 {\rm\,Gyr}$ ago could serve as a building-block
for cluster ellipticals; the halo of our galaxy, considered alone, has
$S_{\rm N}^{\rm P} \simeq 4$.  However, mergers of Milky Way halos (or
dwarf elliptical galaxies \cite{MLFSW98}) still fall short of the high
$S_{\rm N}^{\rm P}$ values of cD galaxies.  Another way to end up with
fewer stars is to eject most of the gas after the initial epoch of
cluster formation; the problem here is that the ejection efficiency
must be {\it higher\/} in cD galaxies, which have the deeper potential
wells and should be better at retaining gas \cite{HHM98}.

Alternately, the production of globular clusters may have been more
efficient in high-redshift starbursts.  Even at low-$z$, about $20$\%
of the UV emitted by starbursts comes from knots identified with young
clusters \cite{MHLKRG95}; if all these clusters survive, the specific
frequency for a pure starburst population is $S_{\rm N} \simeq 60$.
Moreover, these clusters are concentrated where the surface densities
are highest; it's likely that net yields of star clusters increase
rapidly with increasing surface density.

If so, then globular cluster systems reflect the starburst histories
of their parent galaxies: Large populations of metal-poor globulars
are due to efficient cluster production in early starbursts, while
predominantly metal-rich systems (eg., NGC~5846) formed in more recent
starbursts.  Metallicity distributions for cluster systems support
this idea; giant elliptical galaxies have a range of distributions
with multiple peaks between $[{\rm Fe/H}] \simeq -1.2$ and $0.2$
\cite{H94}.  Such variety seems hard to explain in a picture where
internal events determine the timing of cluster formation (eg.,
\cite{FBG97}); on the other hand, it's easy to imagine that different
distributions result from the different merging histories of
individual galaxies.

\section*{CONCLUSIONS}

Circumstantial evidence suggests that merging played an important role
in galactic evolution long before the present epoch.  The key points
of the argument can be summed up as follows:

\begin{enumerate}

\item
Starbursts and AGN are signposts of high-redshift mergers; the high
incidence of such objects at $z \simeq 2$ to~$4$ reflect frequent
merging of juvenile galaxies.

\item
The bulk of the Milky Way's halo merged more than $10 {\rm\,Gyr}$ ago
as part of this activity.

\item
Cluster ellipticals merged before $z \simeq 2$; their immediate
progenitors were few and only moderately gassy.

\item
The metal-rich globular cluster systems of these ellipticals are
relics of their final mergers.

\end{enumerate}

Finally, direct observations of high-redshift events are complemented
by archeological investigation of nearby systems.  Both approaches are
needed to discover what happened at redshifts $z = 2$ to~$5$.

\vspace{0.6 cm}

\noindent
I thank Alex Stephens and Hector Vel\'azquez for communicating results
in advance of publication.  I also thank Jun Makino and the University
of Tokyo for hospitality while I prepared this article.  This research
made use of NASA's Astrophysics Data System Abstract Service.  Travel
to the conference was covered by air miles accumulated while following
the Grateful Dead.

\end{document}